\documentclass[twocolumn]{autart} 
\usepackage[utf8]{inputenc}
\usepackage{bm}
\usepackage{amsfonts}
\usepackage{amsmath}
\usepackage{amssymb}
\usepackage{natbib}
\usepackage{graphicx}
\usepackage{epstopdf}
\usepackage{mathtools}
\usepackage{comment}
\usepackage{xcolor}
\usepackage{empheq}

\newtheorem{remark}{Remark}
\newtheorem{proposition}{Proposition}
\newtheorem{definition}{Definition}
\newtheorem{theorem}{Theorem}
\newtheorem{assumption}{Assumption}

\newcommand{\bs}{\boldsymbol}
\newcommand{\atanTwo}{\operatorname{atan2}}   
\newcommand{\atan}{\operatorname{atan}}
\newcommand{\asin}{\operatorname{asin}}

\newcommand{\ssa}{\operatorname{ssa}}

\begin{document}
\begin{frontmatter} 

\title{A spherical amplitude--phase formulation for 3-D adaptive line-of-sight (ALOS) guidance with USGES stability guarantees}

\thanks[footnoteinfo]{This paper was not presented at any IFAC meeting. Corresponding author E.~M.~Coates. Tel. +47-92669337.}

\author[IIR]{Erlend M. Coates}\ead{erlend.coates@ntnu.no}, 
\author[ITK]{Thor I. Fossen}\ead{thor.fossen@ntnu.no}

\address[IIR]{Department of ICT and Natural Sciences, 
Norwegian University of Science and Technology, {\AA}lesund, Norway}

\address[ITK]{Department of Engineering Cybernetics, 
Norwegian University of Science and Technology, Trondheim, Norway}


\begin{abstract}
A recently proposed 3-D adaptive line-of-sight (ALOS) path-following algorithm addressed coupled motion dynamics of marine craft, aircraft and uncrewed vehicles under environmental disturbances such as wind, waves and ocean currents. Stability analysis established uniform semi-global exponential stability (USGES) using a body-velocity-based amplitude–phase representation of the North–East–Down kinematic differential equations. However, the analysis is limited to straight-line paths, and restrictive assumptions are needed to ensure convergence of the vertical crab angle estimation error to zero. In this paper, we revisit the ALOS framework and introduce a novel spherical amplitude–phase design model that uses an alternative definition of the vertical crab angle. Our proposed formulation enables a significantly simplified stability proof, while retaining the USGES property for straight-line paths, removing restrictive assumptions on constant altitude/depth or zero horizontal crab angle, and remaining valid for general 3-D motion with nonzero roll, pitch and flight-path angles. We also show that the USGES result extends to a class of curved 3-D paths.
\end{abstract}
\end{frontmatter}

\section{Introduction}
Integral (ILOS) and adaptive line-of-sight (ALOS) guidance schemes are well-established path-following algorithms for heading-controlled autonomous vehicles in 2-D~\citep{Borhaug:08,Gu:23,Hung:23,Kjerstad:24}. Strong stability properties, uniform global asymptotic stability (UGAS) and uniform semi-global exponential stability (USGES) are proven, even under environmental drift forces~\citep{Fossen:23TCST}. 3-D extensions for marine craft and aircraft often resort to decoupled planar designs or lack integral action \citep{Lekkas:13,Coates:23}.

Recently, \cite{Fossen:24} proposed a 3-D ALOS law with USGES stability guarantees based on a 3-D amplitude-phase design model. Because it ensures uniform exponential stability in any ball \citep[e.g.,][]{LoriaPanteley:04}, USGES is stronger than UGAS and highly desirable for its robustness properties. However, their stability analysis is limited to straight-line paths and relies on restrictive assumptions (constant depth/altitude or zero horizontal crab angle) for the vertical crab angle estimation error to go to zero.

In this paper, we revisit the framework in \cite{Fossen:24} to address these limitations through the following key contributions: (1) a novel \textit{spherical amplitude--phase representation} of the NED kinematics using a redefined vertical crab angle; (2) a significantly simplified stability proof enabled by this formulation; (3) an extension of the USGES stability result for straight-line path following to fully coupled 3-D motion (nonzero roll, pitch and flight-path angles) without restrictive assumptions; and (4) the extension of these USGES results to "piecewise-planar" curved paths, establishing uniform local exponential stability (ULES) for general 3-D curves. To present these contributions, Section 2 provides kinematic preliminaries, Section 3 introduces the spherical formulation, Section 4 proves USGES for the 3-D ALOS guidance law, Section 5 discusses curved paths and Section 6 concludes the paper.

\section{Kinematic preliminaries}
For clarity of exposition and in line with \cite{Fossen:24}, we consider a piecewise straight-line path defined by a series of waypoints. The extension to curved paths is discussed in Section \ref{sec:curvedpaths}.

The rotation matrix \( \bs{R}_b^n \in SO(3) \) transforms vectors from a BODY frame $\{b\}$, rigidly attached to the vehicle, to a NED reference frame $\{n\}$ and is given by 
\begin{equation}
\bs{R}^n_b =  \left[
\begin{array}
[c]{ccc}%
\mathrm{c}\psi\mathrm{c}\theta & -\mathrm{s}\psi\mathrm{c}\phi+\mathrm{c}%
\psi\mathrm{s}\theta\mathrm{s}\phi & \mathrm{s}\psi\mathrm{s}\phi
+\mathrm{c}\psi\mathrm{c}\phi\mathrm{s}\theta\\
\mathrm{s}\psi\mathrm{c}\theta & \mathrm{c}\psi\mathrm{c}\phi+\mathrm{s}%
\phi\mathrm{s}\theta\mathrm{s}\psi & -\mathrm{c}\psi\mathrm{s}\phi
+\mathrm{s}\theta\mathrm{s}\psi\mathrm{c}\phi\\
-\mathrm{s}\theta & \mathrm{c}\theta\mathrm{s}\phi & \mathrm{c}\theta
\mathrm{c}\phi
\end{array}
\right],  \label{eq:Rxyz}%
\end{equation}
where $\mathrm{s}\;\cdot=\sin(\cdot)$, $\mathrm{c}\;\cdot = \cos(\cdot)$, and $\phi,\theta,\psi$ are the roll, pitch and yaw angles, respectively, following the \( zyx \) Euler angle convention, which is valid for $|\theta|\neq \pi/2$.

During path following, the PATH frame $\{p\}$ is introduced, constructed using two successive rotations: a rotation about the inertial z-axis by the path azimuth angle $\pi_h$, followed by a rotation about the resulting y-axis by the path elevation angle $\pi_v$. For (piecewise) straight-line path following, $\pi_h$ and $\pi_v$ are (piecewise) constant, computed from the path segment between successive waypoints $i$ and $i+1$. For notational convenience, the explicit segment index $i$ is omitted. See \cite{Fossen:24} for details. 

The origin of $\{p\}$ is placed at $\bs{p}^{n}_i = [{x}^n_i, {y}^n_i, {z}^n_i]^\top$, the coordinate vector of the $i$-th waypoint. Further, let $\bs{p}^{n} = [{x}^n, {y}^n, {z}^n]^\top$ be the vehicle's position, expressed in $\{n\}$. 

The \textit{along-}, \textit{cross-} and \textit{vertical-track} errors $x_e^p,  y_e^p,  z_e^p$ expressed in $\{p\}$ are then given by
\begin{equation}
	\bs{e}^p = \left[
	\begin{array}
	[c]{c}%
	x_e^p\\
	y_e^p\\
	z_e^p
	\end{array}
	\right]  =   \bs{R}^{\top}_y(\pi_v)  \bs{R}^{\top}_{z}(\pi_h)  \left(
	\left[  \begin{array}
	[c]{c}%
	x^n\\
	y^n\\
	z^n
	\end{array} \right]
	  - \left[ \begin{array}
	[c]{c}%
	x_i^n\\
	y_i^n\\
	z_i^n
	\end{array} \right] \right).
	\label{eq:error_p}
\end{equation} 
The rotation matrices $\bs{R}_y,\bs{R}_z \in SO(3)$ are elementary rotations about the $y$ and $z$ axes. 

Let $\bs{v}^{b} = [u, v, w]^\top $ be the vehicle's linear velocity vector expressed in $\{b\}$, $U := ||\bs{v}^{b}|| = ||\dot{\bs{p}}^n||$ define the vehicle speed, and $U_h := U \cos(\gamma)$ denote the horizontal speed projection (\emph{speed over ground}). For $U_h \neq 0$, the course (azimuth) angle $\chi \in [-\pi,\pi)$ and the flight-path (elevation) angle $\gamma \in (-\pi/2,\pi/2)$ of the NED velocity vector are then given by
\begin{equation} \label{eq:chigamma}
\chi = \atanTwo(\dot y^n, \dot x^n), \qquad 
\gamma = \asin \left(\frac{- \dot z^n}{U}\right) .
\end{equation}
The NED kinematic differential equation is
\begin{align}
	\dot{\bs{p}}^{n} & =\bs R_b^n \bs{v}^{b}, \label{eq:kinematics}
\end{align}
or equivalently, in terms of spherical coordinates,
\begin{equation} \label{eq:sphericalNED}
\begin{bmatrix}\dot{x}^n \\ \dot{y}^n \\ \dot{z}^n \end{bmatrix} = U \begin{bmatrix} \cos(\gamma)\cos(\chi) \\ \cos(\gamma)\sin(\chi) \\ -\sin(\gamma) \end{bmatrix} .
\end{equation}
The NED kinematic differential equations, \eqref{eq:kinematics}, can be reformulated in amplitude-phase form, which decomposes the velocity into scalar amplitudes and associated angular components, with vehicle orientation as base angles and phase shifts represented by horizontal and vertical crab angles, offsets due to environmental disturbances such as wind, waves and ocean currents.

\subsection{Body-velocity amplitude–phase model} \label{sec:AP1}

The model used in~\cite{Fossen:24}, which we refer to as the \emph{body-velocity amplitude--phase model} since it is defined from body-fixed velocities $u,v,w$, is given as
\begin{align}
    \dot{x}^n  & = U_h^* \cos(\psi + \beta_c^*),  \label{eq:phase-amp-x} \\
    \dot{y}^n  & = U_h^* \sin(\psi + \beta_c^*) , \label{eq:phase-amp-h} \\
	\dot{z}^n  & = -U_v^* \sin(\theta - \alpha_c^*),  \label{eq:phase-amp-v}
\end{align}
where the phase angles, referred to as the horizontal and vertical crab angles, are:
\begin{align}
    \alpha_c^* & = \atan \left( \frac{v \sin(\phi) + w \cos(\phi)}{u} \right),	\label{eq:phase-v} \\
    \beta_c^*  & = \atan \left(\frac{v \cos(\phi) - w \sin(\phi)}{U_v^* \cos(\theta - \alpha_c^*)}\right), \label{eq:phase-h} 
\end{align}
and the amplitudes are given by
\begin{align}
    U_v^*  & = \sqrt{u^2 + \left( v\sin(\phi) + w\cos(\phi) \right)^2 },   \label{eq:speed-v}  \\
    U_h^*  & = \sqrt{\left( U_v^* \cos(\theta - \alpha_c) \right)^2  + \left( v \cos(\phi) - w \sin(\phi) \right)^2}.
 \label{eq:speed-h}
\end{align}
The tracking-error dynamics expressed in $\{p\}$ is found by time differentiation of \eqref{eq:error_p} and substitution of \eqref{eq:phase-amp-x}--\eqref{eq:phase-amp-v}. As shown in~\cite{Fossen:24}, the cross- and vertical-track error dynamics become
\begin{align}
    \dot{y}^p_e  & = U_h^* \sin(\psi + \beta_c^* - \pi_h),  \label{eq:final_ye} \\
    \dot{z}^p_e & = - U_v^* \sin(\theta - \alpha_c^* -\pi_v) + \frac{U_h^*  \sin(\pi_v)} {\sqrt{ 1 + \tan^2(\beta_c^*)}} \notag \\
                & \cdot \left( \sqrt{ 1 + \tan^2(\beta_c^*)  } \, \cos(\psi + \beta_c^* - \pi_h)  - 1  \right) .
	\label{eq:final_ze} 
\end{align}

\section{Spherical amplitude-phase formulation} \label{sec:AP2}
We introduce a novel amplitude--phase representation of the NED kinematic equations using a spherical velocity vector decomposition. 

The orientation of the velocity vector is defined by the course angle $\chi \in [-\pi,\pi) $ and flight-path angle $\gamma \in (-\pi/2,\pi/2)$, defined in \eqref{eq:chigamma}. Similarly, $\psi \in [-\pi,\pi)$ and $\theta \in (-\pi/2,\pi/2)$ define the vehicle's reduced orientation (the azimuth and elevation of the $\{b\}$ x-axis). 
\begin{definition}[Crab angles]
    Let $U_h > 0$ and $|\theta| \neq \pi/2$. The horizontal crab angle $\beta_c \in [-\pi,\pi)$ and vertical crab angle $\alpha_c \in (-\pi,\pi)$ are defined as
\begin{align}
\beta_c := \ssa(\chi - \psi), \qquad \alpha_c := \theta - \gamma .
\label{eq:spherical}
\end{align}
$\ssa(x) := \operatorname{mod}(x + \pi, 2 \pi ) - \pi$ (the smallest signed angle) maps its argument to the principal interval $[-\pi,\pi)$.
\end{definition}
\begin{remark}
    While flight-mechanics literature uses similar expressions using flow angles (sideslip and angle of attack), those only hold individually during restricted, decoupled planar flight. In contrast, our crab angle definitions \eqref{eq:spherical} remain simultaneously valid for any 3-D orientation with $|\theta| \neq \pi/2$ and non-zero horizontal speed $U_h$.
\end{remark}
Substituting \eqref{eq:spherical} into \eqref{eq:sphericalNED} yields the spherical amplitude-phase equations:
\begin{align}
    \dot{x}^n  & = U_h \cos(\psi + \beta_c) , \label{eq:phase-amp-x2} \\
    \dot{y}^n  & = U_h \sin(\psi + \beta_c),  \label{eq:phase-amp-h2} \\
	\dot{z}^n  & = -U \sin(\theta - \alpha_c) \label{eq:phase-amp-v2} .
\end{align}
Here we have used the fact that, since $\cos(\cdot),\sin(\cdot)$ and $\ssa(\cdot)$ are all $2\pi$-periodic, $\cos(\psi + \ssa(\chi-\psi))=\cos(\chi)$ and $\sin(\psi + \ssa(\chi-\psi))=\sin(\chi)$.

Like the body-velocity formulation, the crab angles can be expressed via body-fixed velocities and Euler angles.
\begin{proposition}\label{prop:APF2} 
The phase angles $\alpha_c,\beta_c$ in \eqref{eq:phase-amp-x2}-\eqref{eq:phase-amp-v2} are given by:
\begin{align}
    \alpha_c & = \theta - \asin\left( \frac{u\sin(\theta) - [v \sin(\phi) + w \cos(\phi)]\cos(\theta) }{U} \right),	\label{eq:phase-v2} \\
    \beta_c  & = \atanTwo \bigg( v \cos(\phi) - w \sin(\phi), \notag \\ & \qquad \qquad \,\,\, u\cos(\theta) + [v\sin(\phi) + w\cos(\phi)]\sin(\theta) \bigg) ,\label{eq:phase-h2} 
\end{align}
where $\atanTwo(y,x)$ is the four-quadrant inverse tangent.
\end{proposition}
\vspace{-0.5cm}
\begin{pf}
From trigonometric identities for angle sums, \eqref{eq:phase-amp-x2}-\eqref{eq:phase-amp-h2} expand to
\begin{align}
\dot{x}^n &= -U_h\sin(\beta_c)\sin(\psi) + U_h\cos(\beta_c)\cos(\psi), \nonumber \\
\dot{y}^n &= U_h\cos(\beta_c)\sin(\psi) + U_h \sin(\beta_c)\cos(\psi).\label{eq:xy_dot_AppA}
\end{align}
Comparing this with the first two rows of \eqref{eq:kinematics} gives:
\begin{align}
u\cos(\theta) + [v\sin(\phi) + w\cos(\phi)]\sin(\theta) &= U_h \cos(\beta_c), \notag\\
v\cos(\phi) - w \sin(\phi) &= U_h \sin(\beta_c).\notag
\end{align}
Applying the four-quadrant inverse tangent to these components yields \eqref{eq:phase-h2}:
\begin{align}
\tan(\beta_c) &= \frac{U_h \sin(\beta_c)}{U_h \cos(\beta_c)} = \tfrac{v\cos(\phi) - w \sin(\phi)}{u\cos(\theta) + [v\sin(\phi) + w\cos(\phi)]\sin(\theta) }. 
\end{align}
Comparing the third row of \eqref{eq:kinematics} with \eqref{eq:phase-amp-v2} yields:
\begin{equation}
    -U \sin(\theta - \alpha_c) = -u\sin(\theta) + [v\sin(\phi) + w \cos(\phi)] \cos(\theta) . \notag
\end{equation}
Finally, solving for $\alpha_c$ yields \eqref{eq:phase-v2}.
\end{pf}

By following similar steps as for the body-velocity formulation, the tracking-error dynamics in $\{p\}$ becomes:
\begin{align}
    \dot{y}^p_e  & = U_h \sin(\psi + \beta_c - \pi_h),  \label{eq:final_ye_2} \\
	\dot{z}^p_e &= - U \sin(\theta - \alpha_c -\pi_v) \notag \\
			& + U_h  \sin(\pi_v) \left( \cos(\psi + \beta_c - \pi_h)  - 1  \right). \label{eq:final_ze_2}
\end{align}
Equation \eqref{eq:final_ze_2} is significantly simpler than the original expression \eqref{eq:final_ze}.

\subsection{Comparison with the body-velocity formulation}
\begin{figure}[t]
\centering

\begin{minipage}{\columnwidth}
  \centering
  \includegraphics[width=0.9\linewidth]{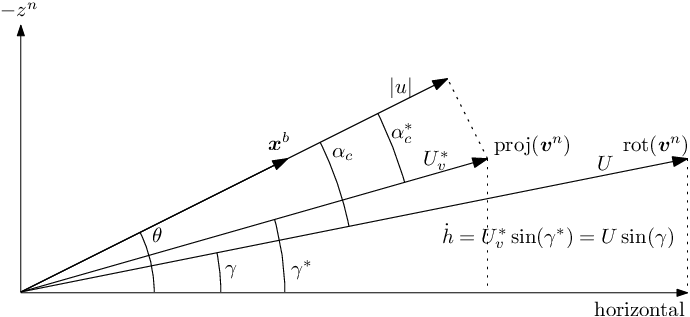}
  \par\vspace{0.5em}  
  \textit{(a) Vertical plane defined by the body $x$-axis.}
\end{minipage}

\vspace{1em}  

\begin{minipage}{\columnwidth}
  \centering
  \includegraphics[width=0.9\linewidth]{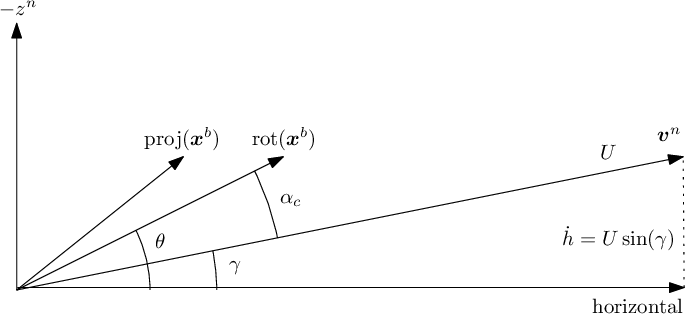}
  \par\vspace{0.5em}
  \textit{(b) Vertical plane defined by the velocity vector.}
\end{minipage}

\vspace{0.5em}
\caption{Illustrating the difference between the formulations.}
\label{fig:vertical_planes}
\end{figure}

The spherical amplitude-phase model parallels \eqref{eq:phase-amp-x}-\eqref{eq:phase-amp-v} but replaces the pair $(U_v^*, \alpha_c^*) $ with $(U, \alpha_c) $. It can be shown directly that the body-velocity form for $\beta_c^*$ in \eqref{eq:phase-h} is equivalent to $\beta_c$ in~\eqref{eq:phase-h2}, and that $U_h = U_h^*$. However, the vertical crab angles are different. The difference is illustrated in Fig.~\ref{fig:vertical_planes}, where the operation $\text{proj}(\cdot)$ projects the vector argument into the plane, while $\text{rot}(\cdot)$ rotates about the inertial z-axis to align with the plane. We see how $\alpha_c^*$ is the angle between the body $x$-axis $\boldsymbol{x}^b$ and the projection of the velocity vector into the vertical plane defined by $\boldsymbol{x}^b$, while $\alpha_c$ is defined in the vertical plane defined by the velocity vector as $\theta-\gamma$. The two are equal only if $\beta_c=0$ or $\gamma=0$, as shown by the relation
\begin{equation} \label{eq:alpha_c_relation}
    \alpha_c^* = \alpha_c + \gamma - \atan\left( \frac{\tan(\gamma)}{\cos(\beta_c)} \right) ,
\end{equation}
which is valid for $\cos(\beta_c) \neq 0$. This singularity occurs when $|\beta_c| = \pi/2$, where the velocity vector's projection into the plane vanishes ($U_v^* = 0$) and $\alpha_c^*$ becomes undefined. In contrast, $\alpha_c$ is globally defined for $U \neq 0$.

\section{ALOS guidance law for 3-D path following} \label{sec:PLOS}
We apply the ALOS guidance law \citep{Fossen:24} to the error dynamics \eqref{eq:final_ye_2}--\eqref{eq:final_ze_2} which is derived using spherical coordinates:
\begin{align}
    \psi_d & =\pi_h - \hat{\beta}_c - \tan ^{-1}\left( \frac{y_e^p}{\Delta_h} \right),  \label{eq:LOS} \\
    \dot{\hat{\beta}}_c & = k_h \frac{\Delta_h}{\sqrt{\Delta_h^2 + (y_e^p)^2}} \, 
\mathrm{Proj}(\hat{\beta}_c,  y_e^p ) , \label{eq:LOS_adap}\\
    \theta_d & = \pi_v + \hat{\alpha}_c + \tan ^{-1}\left( \frac{z_e^p}{\Delta_v}  \right), \label{eq:LOS2} \\
    \dot{\hat{\alpha}}_c & =   k_v \frac{\Delta_v}{\sqrt{\Delta_v^2 + (z_e^p)^2}} \, 
\mathrm{Proj}(\hat{\alpha}_c,  z_e^p ), \label{eq:LOS2_adap}
\end{align}
where $\Delta_h,\Delta_v >0 $ are look-ahead distances, $k_h,k_v>0$ are adaptive gains, and $\hat{\beta}_c,\hat{\alpha}_c$ are crab angle estimates. The projection operator~\citep{Fossen:24} restricts these estimates to compact sets.

\begin{figure}[t]
  \centering
  \includegraphics[width=0.9\linewidth]{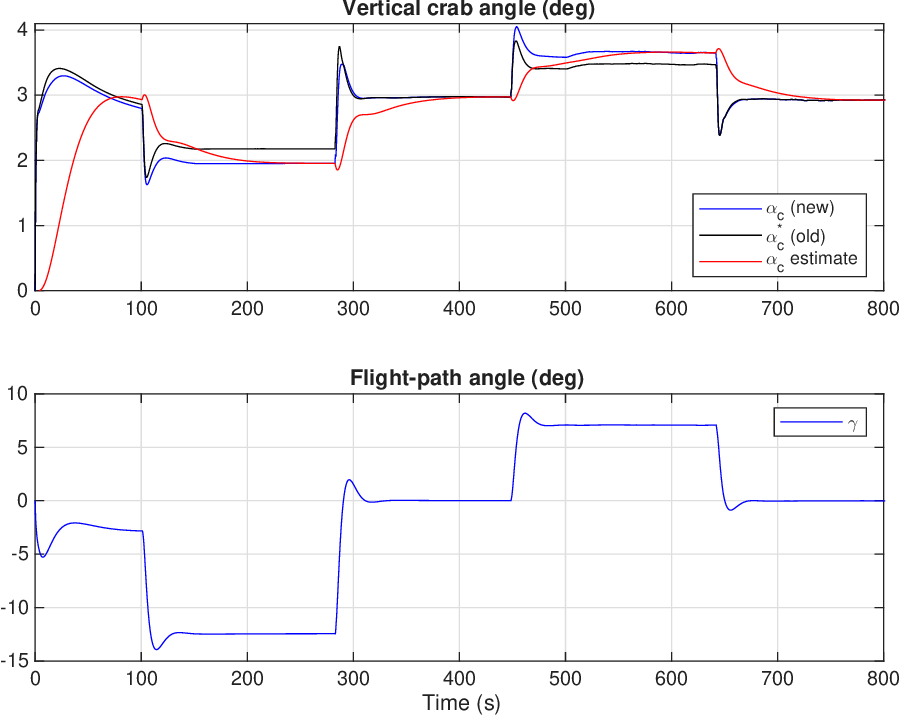}
  \caption{Remus 100 AUV simulation. Vertical bias is removed, also for nonzero horizontal crab and flight-path angles.}
  \label{fig:alpha_comparison}
\end{figure}

The stability analysis relies on the following assumptions, standard in the literature:
\begin{assumption}
    The heading and pitch angle autopilots achieve perfect tracking such that $ \psi = \psi_d$ and $ \theta = \theta_d$.
\end{assumption}
\begin{assumption}
The true crab angles $\alpha_c$ and $\beta_c$ are slowly varying such that $\dot{\alpha}_c \approx 0$ and $\dot{\beta}_c \approx 0$.     
\end{assumption}
Let $\tilde{\alpha}_c = \alpha_c - \hat{\alpha}_c$, $\tilde{\beta}_c = \beta_c - \hat{\beta}_c$. Following \cite{Fossen:24}, inserting \eqref{eq:LOS} and \eqref{eq:LOS2} into \eqref{eq:final_ye_2} and \eqref{eq:final_ze_2}, yields the nonlinear, time-varying cascaded system
\begin{align}				
	&  \hspace{-0.3cm} \Sigma_1: \left\{ 
	\begin{array}{rl}
		\dot{z}_e^p & = -\frac{U \Delta_v}{\sqrt{\Delta_v^2 + (z_e^p)^2}} 
		\left(  \cos(\tilde{\alpha}_c) \frac{z_e^p}{\Delta_v}  -    \sin(\tilde{\alpha}_c) \right)  \\[3pt]
		      & + g(t,y_e^p,\tilde{\beta}_c), \\[5pt]
		\dot{\tilde{\alpha}}_c & = -k_v  \frac{\Delta_v}{\sqrt{\Delta_v^2 + (z_e^p)^2}} 
		\mathrm{Proj}(\hat{\alpha}_c,  z_e^p ), 
	\end{array}		
	\right.  \label{eq:ze_mod} \\
	&  \hspace{-0.3cm} \Sigma_2: \left\{ 
	\begin{array}{rl}
		\dot{y}_e^p & = -\frac{U_h\Delta_h}{\sqrt{\Delta_h^2+ (y_e^p)^2}} 
		\left( \cos(\tilde{\beta}_c)  \frac{y_e^p}{\Delta_h}  -   \sin(\tilde{\beta}_c) \right),  \\[5pt] 
  		\dot{\tilde{\beta}}_c & = -k_h \frac{\Delta_h}{\sqrt{\Delta_h^2 + (y_e^p)^2}} 
  		\mathrm{Proj}(\hat{\beta}_c,  y_e^p ) ,
	\end{array}
	\right. \label{eq:ye_mod} 
\end{align}
\begin{equation}
	g(t,y_e^p,\tilde{\beta}_c) \! = \! U_h  \sin(\pi_v) \left(  \, \cos \left( \tilde{\beta}_c \!
     	- \! \atan\left( \frac{y_{e}^p}{\Delta_h} \right) \right) \! - \!  1  \right) .	\label{eq:ddef2} 
\end{equation}
Although $\Sigma_1$--$\Sigma_2$ resembles the closed-loop system in \cite{Fossen:24}, the redefined coupling term, \eqref{eq:ddef2}, alters the analysis, enabling a simplified, and crucially, less restrictive stability result in Theorem 1.
\begin{theorem}[USGES of the ALOS guidance law] \label{theorem1}
Under assumptions 1-2 and bounded vehicle speed ($0 < U_{\min} \leq U(t) \leq U_{\max}$), the ALOS guidance law \eqref{eq:LOS}--\eqref{eq:LOS2_adap}, with parameters $\Delta_h,\Delta_v,k_h,k_v> 0$ applied to the cross- and vertical-track errors \eqref{eq:final_ye_2}--\eqref{eq:final_ze_2}, renders the origin $(z_e^p,\tilde{\alpha}_c,y_e^p,\tilde{\beta}_c) = (0,0,0,0)$ of $\Sigma_1$--$\Sigma_2$ USGES.
\end{theorem}
\vspace{-0.3cm}
\begin{pf}
From \cite{Fossen:24}, the origin $(z_e^p,  \: \tilde{\alpha}_c) = (0, \,  0)$ of $\Sigma_1$ is USGES when the perturbation $ g(t,y_e^p,\tilde{\beta}_c) \equiv 0$. Furthermore, the origin $(y_e^p,  \: \tilde{\beta}_c) = (0, \,  0)$ of $\Sigma_2$ is USGES. Together, \eqref{eq:ze_mod}--\eqref{eq:ye_mod} form a cascade. Since $0 < U_{\min} \leq U \leq U_{\max}$, the perturbation term \eqref{eq:ddef2} is bounded and $ g(t,0,0) = 0$. By \citet[Lemma 9.4, Corollary 1]{Khalil:02}, the perturbed $\Sigma_1$ states are uniformly ultimately bounded. Since $\Sigma_2$ is USGES, $y_e^p$ and $\tilde{\beta}_c$ converge exponentially to zero. Consequently, the perturbation term $g(t,y_e^p,\tilde{\beta}_c)$ vanishes, and we conclude that the equilibrium of the cascade $\Sigma_1$--$\Sigma_2$ is USGES.
\end{pf}
\vspace{-0.3cm}
\begin{remark}
Unlike \cite{Fossen:24}, which assumes $\beta_c = 0$ or $\pi_v = 0$ for vertical crab-angle estimate convergence, our formulation achieves USGES without these restrictions. The vertical perturbation \eqref{eq:ddef2} vanishes, independent of $\beta_c,\pi_v$, guaranteeing horizontal and vertical error convergence during coupled 3-D motion.
\end{remark}
This updated result eliminates steady-state vertical bias ($\hat{\alpha}_c \to \alpha_c$) under a broader range of conditions, as illustrated in Fig.~\ref{fig:alpha_comparison}, where we reproduce the Remus 100 AUV simulation from \cite{Fossen:24} using modified gains $k_h = k_v = 0.0015$ (originally denoted $\gamma_h, \gamma_v$). The estimate converges to $\alpha_c$ rather than $\alpha_c^*$, validating our redefined vertical crab angle. Furthermore, the presence of a horizontal current ensures $\beta_c \neq 0$; thus, consistent with \eqref{eq:alpha_c_relation}, $\alpha_c = \alpha_c^*$ occurs only when $\gamma = 0$ (bottom subplot). 

\section{Extension to Curved Paths} \label{sec:curvedpaths}
\vspace{-0.3cm}
We briefly sketch the extension of Theorem 1 to regular parametrized curved paths \citep{Breivik:05}. Here, $\pi_h(\varpi),\pi_v(\varpi)$, parametrize the path tangent vector and vary as a function of the path-parameter $\varpi$. Differentiating \eqref{eq:error_p} now gives $- \bs{\omega}_p^p \times \bs{e}^p$, where the angular velocity $\bs{\omega}_p^p = [\omega_x,\omega_y,\omega_z]^\top$ depends on path curvature and $\dot{\varpi}$. Selecting $\varpi$ to maintain identically zero along-track error $x_e^p$ \citep{Schmidt-Didlaukies:24} yields
\begin{align}
    \dot{y}^p_e  & = U_h \sin(\psi + \beta_c - \pi_h) + \omega_x z_e^p,  \\
	\dot{z}^p_e &= - U \sin(\theta - \alpha_c -\pi_v) - \omega_x y_e^p \notag \\
			& + U_h  \sin(\pi_v) \left( \cos(\psi + \beta_c - \pi_h)  - 1  \right).
\end{align}
For our frame $\{p\}$ (horizontal y-axis), $\omega_x = -\sin(\pi_v)\dot{\pi}_h$. We thus conclude that Theorem 1 generalizes to curved paths that are piecewise either horizontal ($\pi_v = 0$) or contrained to the same vertical plane ($\dot{\pi}_h=0$). In this case, $\omega_x = 0$, recovering \eqref{eq:final_ye_2}-\eqref{eq:final_ze_2}, satisfying Theorem 1.
For general 3-D paths (bounded $\omega_x \neq 0$), a linear-growth vanishing perturbation perturbs the nominal (straight-line) dynamics. Let $\bs{\xi}=[z_e^p,\tilde{\alpha}_c,y_e^p,\tilde{\beta}_c]^\top$. By the USGES property and converse Lyapunov theorem~\citep{Khalil:02}, the unperturbed system admits a quadratic strict Lyapunov function $V(\bs{\xi})$ for any $r>0$ and $||\bs{\xi}||< r$, dominating the perturbation for small $|\omega_x|$. Since the maximum allowable $|\omega_x|$ depends on $r$, we conclude only uniform local exponential stability (ULES) for $\bs{\xi}=0$.

\vspace{-0.3cm}
\section{Conclusions}
\vspace{-0.3cm}
This paper introduces a novel spherical amplitude--phase representation of NED kinematic equations for 3-D path following, utilizing a redefined vertical crab angle. This formulation avoids geometric singularities, simplifies error dynamics, and eliminates steady-state bias in vertical crab-angle estimates. It guarantees USGES with a simpler proof and less restrictive assumptions, valid for general 3-D motion, accommodating nonzero roll, pitch, horizontal crab and flight-path angles. While these results extend to a class of curved paths, general 3-D curves yield ULES. Achieving global stability results for arbitrary curved paths remains a topic for future work.
\vspace{-0.3cm}


\end{document}